\documentclass[print,twocolumn]{revtex4}
\usepackage{amssymb}
\usepackage{amsmath}
\usepackage{graphicx}
\usepackage{dcolumn}
\usepackage{bm}
\usepackage{txfonts}
\usepackage{appendix}
\usepackage[dvips]{color}
\usepackage[colorlinks, linkcolor=blue, anchorcolor=blue, citecolor=blue, urlcolor=blue ]{hyperref}

\begin{document}

\title{Power law decay of local density of states oscillations near a line defect in a system
with semi$-$Dirac points}
\author{Wang Chen}
\author{Xianzhe Zhu}
\author{Xiaoying Zhou}
\email{xiaoyingzhou@hunnu.edu.cn}
\author{Guanghui Zhou}
\affiliation{Department of Physics, Key Laboratory for Low-Dimensional Structures and
Quantum Manipulation (Ministry of Education), Hunan Normal University,
Changsha 410081, China}

\begin{abstract}
We theoretically study the power-law decay behavior of the local density of
states (LDOS) oscillations near a line defect in system with semi-Dirac
points by using a low-energy $k$$\cdot$$p$ Hamiltonian. We find that
the LDOS oscillations are strongly
anisotropic and sensitively depend on the orientation of the line defect. We
analytically obtain the decay indexes of the LDOS oscillations near a line defect
running along different directions by using the stationary phase
approximation. Specifically, when the line defect is perpendicular to the
linear dispersion direction, the decay index is $-5/4$ whereas it becomes $-1/4$
if the system is gapped, both of which are different from the decay index $-3/2$ in isotropic
Dirac systems. In contrast, when the line defect is perpendicular to the
parabolic dispersion direction, the decay index is always $-1/2$ regardless of
whether the system is gapped or not, which is the same as that in a
conventional semimetal. In general, when the defect runs along an arbitrary
direction, the decay index
sensitively depends on the incident energy for a certain orientation of the line defect. It varies from $-5/4$ to $-1/2$ due to
the absence of strict stationary phase point. Our
results indicate that the decay index $-5/4$ provides a fingerprint to
identify semi-Dirac points in 2D electron systems.
\end{abstract}

\maketitle


\section{Introduction}

The discovery of graphene has triggered a boom of study on the Dirac-Weyl
fermions in condensed matter systems on account of both rich physics therein
and promising applications \cite{Novoselov,Neto}. Graphene possesses a
gapless energy spectrum with linear dispersion around two inequivalent Dirac
points in its Brillouin zone \cite{Neto}. This peculiar band structure
contributes to unique transport properties such as Klein tunneling \cite{Katnelson} and half-integer quantum-Hall effect \cite{Zhang1}. When a
graphene is subjected to anisotropic strain, the nearest hoppings also
become anisotropic, and the Dirac points will move towards each other \cite{Dietl,Montambaux1,Sinha,Hasegawa,Pereira}. Under critical anisotropy, two inequivalent Dirac points
merge into a semi-Dirac point (SDP), around which the energy dispersion is
linear in one direction and parabolic in the perpendicular direction \cite{Dietl,Montambaux1,Sinha}.

Besides strained graphene, SDP in energy spectrum has also been predicted in
many other systems, such as the strained or electric field modulated
few-layer black phosphorus \cite{Kim,Baik,Wang1,Liu,Ghosh,Doh}, multilayer $%
\mathrm{(TiO_{2})_{n}/(VO_{2})_{ m}}$ nanostructures \cite%
{Pardo1,Banerjee1,Huang}, silicene oxide \cite{Zhong}, Bi$_{1-x}$Sb$%
_{x}$ thin film \cite{Tang}, striped boron sheet \cite{Zhang2}, strained
monolayer arsenene \cite{Wang2} and spiral multiferroic oxide modulated
surface states in topological insulators \cite{Li,Zhai1}. To date,
semi-Dirac spectrum has been observed experimentally in potassium doped
few-layer black phosphorus \cite{Kim}, tunable ultracold
atomic honeycomb optical lattice \cite{Tarruell}, and polariton honeycomb lattices \cite{Real}. Although the dispersion around a SDP is a combination of that in conventional semimetals and
Dirac materials, the low-energy physics in it may exhibit unique features
which can't be fully understood by combing the existed results such
as the unusual Landau levels \cite%
{Dietl,Banerjee1,Montambaux1,Delplace}, optical conductivity
\cite{Jang,Carbotte}, anisotropic plasmon \cite{Banerjee2}, and Fano factor
in ballistic transport \cite{Zhai2}.

Impurities and defects in materials induce many interesting physical phenomena, such as the quasi-particle interference (QPI) patterns \cite{Friedel,Crommie,kehuiwu,Avraham} and the RKKY interaction between magnetic defect lines \cite{Gorman} in graphene. QPI induced by line defects or point
impurities gives an oscillation pattern of the local
density of states (LDOS) in the vicinity of the imperfections \cite{Friedel,Crommie,kehuiwu,Avraham}. Those LDOS oscillations can be directly probed using the scanning tunneling microscope \cite{Petersen,Crommie,kehuiwu,Avraham,Xue}. The wave
vector corresponding to the QPI pattern depends on the geometry of the
constant energy contour (CEC) \cite{Petersen}. Hence, the relevant properties of Fermi
surface can be extracted from the LDOS, which makes the QPI image is
particularly useful in probing the dispersion of the surface bands \cite{Petersen,Crommie,kehuiwu,Avraham,Xue}. In turn,
the QPI patterns exhibit unique characteristics in different electron
systems \cite{Crommie,Xue,Wang}. In system with isotropic Dirac points such as graphene or the surface states of
three dimensional (3D) topological insulators, the power-law decay behavior of
LDOS near a line defect is $x^{-3/2}$ \cite{Xue,Wang}, which
is much faster than $x^{-1/2}$ in conventional two-dimensional (2D) electron
gas \cite{Crommie}, where $x$ is the distance away
from the line defect. These decay indexes serve as fingerprints to characterize related
physical systems \cite{Crommie,Xue,Wang}. Since the low-energy dispersion around SDP is inherited
from that in conventional semimetals and isotropic Dirac materials, a natural
question is what is the power-law decay index of the LDOS oscillations near a line
defect in electron system with SDPs$?$

Herein, this work studies the power-law decay behavior of the LDOS oscillations near a
line defect in a 2D semi-Dirac system. The line defect
is modeled by an ultrathin high rectangular barrier, which is also
adopted in previous works \cite{Biswas1,Biswas2,Wang} studying the LDOS oscillation near it on the surface of 3D topological insulators.
Using a low-energy $k $$\cdot$$p$ Hamiltonian, we find
that the LDOS oscillations are strongly anisotropic, sensitively depending
on the orientation of the defect. We analytically obtain the decay indexes of the LDOS oscillations in various cases
by using stationary phase approximation \cite{Biswas1,Biswas2,Wang,Liuq}. Specifically, when the line defect is
perpendicular to the linear dispersion direction, the decay index of the
LDOS is $-5/4$ whereas the it becomes $-1/4$ if the SDP is gapped, both of which
are different from the index $-3/2$ in system with isotropic Dirac points. However,
when the line defect is perpendicular to the parabolic dispersion direction,
the decay indexes are $-1/2$ regardless of the SDP is gapped or not,
which is the same as that in a conventional semimetal. Further, when the line defect is perpendicular
to an arbitrary direction between the linear and parabolic dispersion
directions, the
decay index sensitively depends on the incident energy due to the absence of strict stationary phase point. It varies from $-5/4$ to $-1/2$. Our results indicate that the
power-law decay index $-5/4$ provides a fingerprint to verify the SDP in 2D
electron systems.

The rest of this paper is organized as follows. In Sec. II, we introduce the
low-energy effective model and the stationary phase approximation. Sec. III
presents some numerical results and discusses of the LDOS oscillations in various cases
combined with analytical analysis based on the stationary phase
approximation. In Sec. IV, we summarize our work.

\section{Model and Method}
The effective low-energy Hamiltonian around a semi-Dirac point is \cite{Dietl,Baik}
\begin{equation}
H=\frac{\hbar ^{2}k_{y}^{2}}{2m^{\ast }}\sigma _{x}+\hbar v_{F}k_{x}\sigma
_{y}+\Delta \sigma _{z},
\end{equation}%
where $\sigma _{x}$, $\sigma _{y}$ and $\sigma _{z}$ are the Pauli matrices,
$m^{\ast }$ the effective mass and $v_{F}$ the Fermi velocity, and $\mathbf{k%
}=(k_{x},k_{y})$ the wavevector. We also include a gap $\Delta$ in
Hamiltonian (1) to explore whether it impacts the LDOS oscillation or not.
Typically, the two parameters in potassium doped few-layer black phosphorus
\cite{Baik} are $v_{F}=3\times 10^{5}$ m/s and $m^{\ast }=1.42$ $m_{e}$,
where $m_{e}$ is the free electron mass. According to the data of the angular resolved photoelectron spectroscopy measurement in Ref \cite{Kim}, Hamiltonian (1) is valid in the energy regime within 0.4 eV relative to the semi-Dirac point. The corresponding eigenvalue is
\begin{equation}
E_{\pm }(k_x,k_y)=\pm \sqrt{\frac{\hbar ^{4}k_{y}^{4}}{4m^{\ast 2}}+\hbar
^{2}v_{F}^{2}k_{x}^{2}+\Delta ^{2}},
\end{equation}%
where $E_{+/-}$ is the conduction/valence band. The eigenvector is
\begin{equation}
\Psi_{\mathbf{k}}(\mathbf{\ r}) =\left(%
\begin{array}{c}
1 \\
\chi%
\end{array}%
\right) e^{i\mathbf{k\cdot r}},
\end{equation}%
with $\chi =(\hbar ^{2}k_{y}^{2}/2m^{\ast }+i\hbar v_{F}k_{x})/(E+\Delta )$.
For $\Delta =0$, Eq. (2) is the energy dispersion near a SDP. As
shown in Fig. 1(c), the energy band linearly (parabolically) disperses
along the $k_{x}$ ($k_y$) direction. Fig. 1(d) depicts the density of states
(DOS) corresponding to Fig. 1(c). In contrast to the linearly dependent DOS
around the isotropic Dirac point \cite{Neto}, the DOS around the SDP is
proportional to $E^{1/2}$ \cite{Banerjee1,Banerjee2}.
\begin{figure}[tbp]
\center
\includegraphics[bb=0 0 418 405, width=3.4 in]{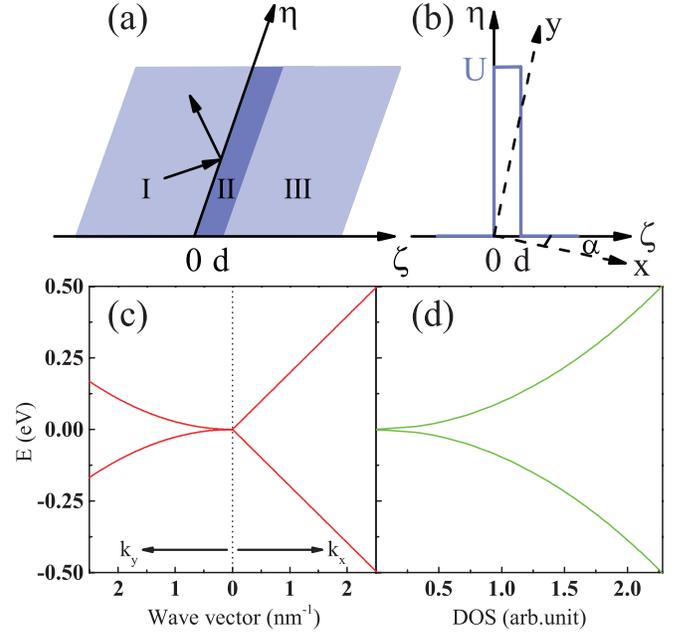} \renewcommand{%
\figurename}{FIG.}
\caption{(a) Schematic diagram of the scattering problem. (b) The profile of the electric potential. (c) Band structure
and (d) density of states near a 2D semi-Dirac point.}
\end{figure}

Following previous works studying the LDOS oscillations near a line defect on the surface of 3D topological insulators \cite{Biswas1,Biswas2,Wang}, we model the line defect as a ultrathin rectangular electric barrier
[see Figs. 1(b)]. The limitation of this barrier is a $\delta$-potential if we
keep $Ud\equiv$ constant with decreasing width ($d$$\rightarrow$0). We also study the LDOS oscillations near the line defect by modeling it as a $\delta$-potential in the Appendix.
Without loss of generality, we assume that the barrier is parallel (perpendicular) to the $\eta$ ($\zeta$) direction at angle $\alpha \in [0, \pi/2]$ with respect to the $x$-axis [see Fig. 1(b)]. Where $\alpha=0$/$\frac{\pi}{2}$ corresponds to the line
defect perpendicular to the linear/parabolic dispersion direction. The
potential profile is $U(\zeta)=U_{0}[\Theta(\zeta)-\Theta(\zeta-d)]$ with $%
\Theta(\cdot)$ the Heaviside step function. Then, the wave vectors $k_x$ and
$k_y$ can be transformed in terms of $k_{\zeta}$ and $k_{\eta}$ \cite%
{Banerjee2}, i.e., $k_{x}=k_{\zeta}\cos\alpha-k_{\eta} \sin\alpha$ and $%
k_{y}=k_{\zeta}\sin\alpha+k_{\eta}\cos\alpha$.

As shown in Fig. 1(a), the scattering frame is divided into three regions,
i.e., the incident region I, the barrier region II, and the transmitted
region III. Owing to the translation invariance in the $\eta$-direction, the
transverse wave vector $k_{\eta}$ is a good quantum number. Therefore, the
wave function admits the form $\Psi_{\mathbf{k}}(\zeta,\eta)=e^{ik_{\eta
}\eta}\varphi(\zeta)$. The scattered wave is characterized by the
longitudinal wave vector $k_{\zeta}$ result from the energy conservation. For briefness, we express all quantities in
dimensionless units by introducing a length unit $l_0=1$ nm and an energy
unit $E_{0}=\hbar^{2}/(2m^{\ast}l_0^{2})=26.8$ meV. Hereafter, the lengths
(energies) are in unit of $l_0$ ($E_0$), and the wavevectors are in unit of 1/$%
l_0$ throughout the paper. For certain $k_{\eta}$ and $E$, the longitudinal
wave vectors $k_{\zeta}$ of each region are governed by
\begin{equation}
\left(E-U\right)^{2}=\Delta^{2}+[h_{1}(k_{\zeta},k_{\eta})]^{4}+u^{2}[h_{2}(k_{\zeta},k_{\eta})]^{2},
\end{equation}
where $h_{1}(k_{\zeta},k_{\eta})=k_{\zeta}\sin\alpha+k_{\eta}\cos\alpha$, $h_{2}(k_{\zeta},k_{\eta})=k_{\zeta}\cos\alpha-k_{\eta} \sin\alpha$, and the dimensionless quantities $u=2m^{\ast}v_{F}l_0/\hbar$. This is a quartic algebraic equation about the wave vector $k_{\zeta}$ except for the case of $\alpha=0$. For $\alpha\neq0$, the wave function in region $j$ $(j=%
\mathrm{{I,II,III})}$ is
\begin{equation}
\begin{aligned} \Psi_{j}=&\left[a_{j_{1}}\left(\begin{array}{c} 1 \\
\chi_{j_{1}}\end{array}\right)e^{ik_{\zeta j_{1}}\xi
}+a_{j_{2}}\left(\begin{array}{c} 1 \\
\chi_{j_{2}}\end{array}\right)e^{ik_{\zeta j_{2}}\xi }\right. \\&
\left.+a_{j_{3}}\left(\begin{array}{c} 1 \\ \chi_{j_{3}}\end{array}\right)
e^{ik_{\zeta j_{3}}\xi }+a_{j_{4}}\left(\begin{array}{c} 1 \\
\chi_{j_{4}}\end{array}\right)e^{ik_{\zeta
j_{4}}\xi}\right]e^{ik_{\eta}\eta}, \end{aligned}
\end{equation}
where $k_{\zeta_{n}}$ $(n=1,2,3,4)$ are the four solutions of Eq. (4). In
region I $(\zeta<0)$, an incident mode $k_{\zeta }^{i}=k_{\zeta_{1}}$ propagating to the
right with $v_{\zeta}=({\partial}E/{\partial}k_{\zeta})_{k_{\eta}}>0$ may be
scattered into a reflected mode $k_{\zeta }^{f}=k_{\zeta_{2}}$ propagating towards the left
with $v_{\zeta}<0$ and an evanescent mode with Im$(k_{\zeta_{3}})<0$ . In
region II $(0\leqslant \zeta \leqslant d)$ , four modes exist due to the
ultrathin barrier. In region III $(\zeta>d)$, there is a transmitted mode
propagating to the right and an evanescent mode with Im$(k_{\zeta_{4}}) >0$.
Based on the above analysis, we have $a_{\mathrm{I_{4}}}=a_{\mathrm{III_{2}}%
}=a_{\mathrm{III_{3}}}=0$. We also set $a_{\mathrm{I_{1}}}=1$ in region I to
simplify the calculations. The rest eight unknown coefficients can be
determined by applying the boundary conditions of the wave functions and
probability current, which are given by
\begin{equation}
\begin{aligned}
\Psi_{\mathrm{I}}|_{\zeta=0}&=\Psi_{\mathrm{II}}|_{\zeta=0},\quad
\Psi_{\mathrm{II}}|_{\zeta=d}=\Psi_{\mathrm{III}}|_{\zeta=d}, \\
\hat{v}_{\zeta}\Psi _{\mathrm{I}}|_{\zeta=0}&=\hat{v}_{\zeta}\Psi
_{\mathrm{II}}|_{\zeta=0}, \quad \hat{v}_{\zeta}\Psi
_{\mathrm{II}}|_{\zeta=d}= \hat{v} _{\zeta}\Psi_{\mathrm{III}}|_{\zeta=d},
\end{aligned}
\end{equation}
where $\hat{v}_{\zeta}$=$\partial\hat{H}/\partial k_{\zeta}$ is the current
operator. Then, the unknown coefficients such as the reflection
amplitudes $r$=$a_{\mathrm{I_{2}}}$ can be determined by using the transfer
matrix method \cite{Dignam,Sedrakian} combined with the boundary conditions
in Eq. (6). For $\alpha=0$, Eq. (4) is a quadratic equation about $k_{x}$.
There are only two real solutions for $k_{x}$, which means there is no
evanescent mode in Eq. (5). Similarly, we can set $a_{\mathrm{I_{1}}}$=$1$,
and the rest four nonzero coefficients can be solved using only the boundary
conditions of the wave functions, i.e., $\Psi_{\mathrm{I}}|_{x=0}=\Psi _{%
\mathrm{II}}|_{x=0}$, and $\Psi_{\mathrm{II}}|_{x=d}=\Psi_{\mathrm{III}%
}|_{x=d}$.

In region I, the interference between the incident and reflected waves gives
an oscillation pattern of the LDOS in real space, i.e., the Friedel
oscillations \cite{Friedel}. The LDOS
near a line defect is \cite{Biswas1,Biswas2,Wang,Liuq}
\begin{equation}
\begin{aligned}
\rho (\zeta ,E) &=\sum\limits_{\mathbf{k}}\left\vert \Psi _{1}\left( \zeta
,\eta \right) \right\vert ^{2}\delta (E-E_{\mathbf{k}}) \\
&=\int \left\vert \Psi _{1}\left( \zeta ,\eta
\right) \right\vert ^{2}\delta (E-E_{\mathbf{k}})d{\mathbf{k}} \\
&=\oint\limits_{E}\left\vert \Psi _{1}\left(
\zeta ,\eta \right) \right\vert ^{2}dk_{\eta }=\rho
_{0}\left( E\right) +\delta \rho \left( \zeta ,E\right),
\end{aligned}
\end{equation}%
where $\rho _{0}(E)$ is spatially independent, and it can be ignored. In real
scanning tunneling microscope experiments \cite{Petersen,Crommie,kehuiwu,Avraham,Xue}, one often measures the spatially dependent part $\delta \rho(\zeta ,E)$, which is given by
\begin{equation}
\begin{aligned}
\delta \rho \left( \zeta ,E\right)=\oint_{E}\mathrm{Re}\left[
r\left( 1+\chi _{\mathrm{I}_{1}}^{\ast }\chi _{\mathrm{I}_{2}}\right)
e^{i\left( k_{\zeta }^{f}-k_{\zeta }^{i}\right) \zeta }%
\right] dk_{\eta }+\delta \rho _{1}\left( \zeta ,E\right),
\end{aligned}
\end{equation}%
where $\delta \rho _{1}\left( \zeta ,E\right) $ originates from the
evanescent mode and decays to zero quickly for positions far away from the line defect. Therefore, the LDOS oscillation is dominated by the
first term in Eq. (8). For positions away from the defect, the LDOS
oscillation sensitively depends on the phase factor $e^{i(k_{\zeta }^{f}%
-k_{\zeta }^{i})\zeta }$ which oscillates rapidly. A pair of
scattering sates ($k_{\zeta }^{i},k_{\eta }$) and ($k_{\zeta }^{f},k_{\eta }$%
) on the CEC result in a standing wave with spatial period of $2\pi
/|(k_{\zeta }^{f}-k_{\zeta }^{i})|\equiv 2\pi /|\Delta k_{\zeta }|$. Only the pair whose period is stationary with respect to small variation in $%
k_{\eta }$ makes dominant contribution to the LDOS oscillations \cite{Wang,Biswas2,Liuq}. Those pairs of points on the CEC are called as
stationary phase points, which satisfy
\begin{equation}
\frac{\partial \Delta k_{\zeta }}{\partial k_{\eta }}|_{k_{\eta0}}=\frac{%
\partial \left( k_{\zeta }^{f}-k_{\zeta }^{i}\right) }{\partial k_{\eta}}%
|_{k_{\eta0}}=0.
\end{equation}%
The stationary phase points given by Eq. (9) can be divided into two categories according to the sign of the second derivative ($\partial^2 \Delta k_\zeta/\partial k_\eta^2$) in the neighbourhood of these points. One category is the extreme points (EPs) around which the second derivatives have the same signs \cite{Wang,Biswas2,Liuq}.  In this case, the wavevector changes $\Delta k_\zeta$ are maximum or minimum values on the CEC. Another category is the inflection points around which the second derivatives have opposite signs. In our work, we only encounter the EPs. Between a pair of EPs, $\Delta k_{\zeta0}=k_{\zeta0}^{f}-k_{\zeta0}^{i}$ is the characteristic wavevector solely determined by the geometry of CEC. The spatial dependence of the LDOS can be evaluated by expanding the relevant quantities in Eq. $(8)$ to the
lowest leading order about $\delta k_{\eta }$ around each pair of EPs, which is
given by
\begin{equation}
\begin{aligned} k_{\eta}&\rightarrow k_{\eta0}+\delta k_{\eta }, \quad
r\rightarrow r_{0}\delta k_{\eta }^{\beta_{\zeta} }, \\ 1+\chi
_{\mathrm{I}_{1}}^{\ast }\chi _{\mathrm{I}_{2}}&\rightarrow c_{0}\delta
k_{\eta }^{\gamma_{\zeta} }, \quad k_{\zeta }^{f}-k_{\zeta }^{i}\rightarrow\Delta k_{\zeta0}+\Delta
k_{\zeta\lambda }\delta k_{\zeta }^{\lambda_{\zeta} }, \end{aligned}
\end{equation}%
Then, the asymptotic behavior of the LDOS is
\begin{equation}
\delta \rho (\zeta ,E)\simeq\rho_A \cos (\Delta k_{\zeta0}\zeta +\phi )\zeta
^{\nu },
\end{equation}%
where $\rho_A=\mathrm{Re}(i^{-\nu}c_0r_0/\lambda_{\zeta})\Gamma(-\nu)(\Delta k_{\zeta\lambda})^{\nu}$ is the amplitude of LDOS pattern, $\Delta k_{\zeta\lambda}$ is the lowest leading order derivative of $\Delta k_{\zeta}(k_{\eta})$ at ($k_{\xi}^i,k_{\eta}$), $\nu =-(\beta _{\zeta }+\gamma _{\zeta
}+1)/\lambda _{\zeta }$ is the power-law decay index, $\Gamma(x)$ is the Euler function, and $\phi $ is the initial phase of each pair of EPs. The asymptotic decay behavior of LDOS oscillations in Eq. (11) is valid if $\zeta\gg|\Delta k_{\zeta0}|^{-1}$ \cite{Wang,Biswas2,Liuq}, which means the asymptotic region is energy dependent.

\section{Local density of states oscillations}

In this section, we present some numerical examples of the LDOS oscillations
when the line defects are along different directions for the system with and
without a gap, respectively. In order to understand the numerical results
better, we analytically obtain the power-law decay indexes of the LDOS
oscillations for two special cases within the stationary phase
approximation.


\begin{figure}[t]
\center
\includegraphics[bb=181 30 572 583, width=3.4 in]{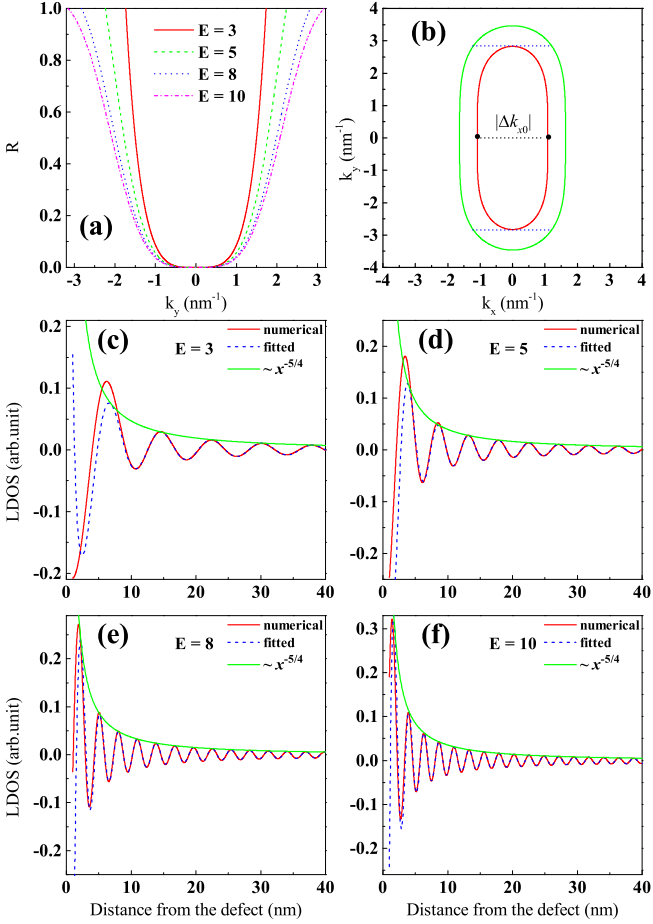} %
\renewcommand{\figurename}{FIG.}
\caption{(a) The reflectivity $R$ as a function of $k_y$ at different energies for $\protect\alpha=0$. (b) The CECs of
the incident (red line) and scattering (green line) regions with $E=8$. The
black solid dots are a pair of stationary phase points, and $|\Delta
k_{x0}|$ is the characteristic wave vector of the scattering process.
(c)-(f) The spatial dependence of LDOS at different energies. The red solid (blue dashed)
lines are the numerical (fitted) results. The green solid lines are the envelop function of
LDOS. In all figures, we have set $%
U_0=20$ and $d=1$.}
\end{figure}

First, we consider the LDOS oscillation of $\alpha =0$ for the gapless case (%
$\Delta =0$). Fig. 2(a) plots the reflectivity $R=|r|^{2}$ as a function of
the transverse wave vector $k_y$ with different energies. As shown in the figure, the
reflectivity $R$ is always zero at normal incidence i.e., $k_y=0$, due to the Klein
tunneling resulting from the time-reversal symmetry of
the effective Hamiltonian \cite{Banerjee2,Jiang}. It increases at oblique incidence with increasing incident
angle due to the mismatch of the wave vector \cite{Banerjee2,Zhai2}. Fig.
2(b) depicts the CECs of the incident area (red line) and the barrier area
(green line). As plotted in the figure, the scattering only occurs
in the range within the blue dashed lines resulting from the conservation of $%
k_{y}$. Figs. 2(c-f) present the spatial dependence of LDOS with $E=3,5,8,10$%
, respectively. As depicted in the figure, numerical results (the solid red
lines) indicate that the LDOS periodically oscillates with the distance away
from the defect with a decreasing amplitude, which implies an asymptotic
behavior.

As discussed in Sec II, the LDOS oscillation can be
understood by using the stationary phase approximation. First, by
using Eq. (9), the pair of stationary phase points on the CEC are $(E/u,0)$ and $%
(-E/u,0)$ [see the black solid dots in Fig. 2(b)]. The second derivative ($\partial^2 \Delta k_x/\partial k_y^2$) at these points are zero but negative in the neighbourhood of them. Hence, these two points $[(\pm E/u,0)$] constitute a pair of maximum points, giving the characteristic
wave vector as $|\Delta k_{x0}|=|-2E/u|$. Therefore, the period of the LDOS
oscillation is $2\pi /|\Delta k_{x0}|=\pi u/E$, which means the higher the incident energy, the faster the LDOS oscillations.
This explains why the LDOS oscillates faster for higher energy in Figs.
2(c-f). In order to get the power-law decay index, we need to expand relevant quantities around the EPs. Because the first, second, and third derivatives of $\Delta k_x(k_y)$ at the EPs are all zero, we have to expand it to the fourth-order. Then, near the EPs, we have $\Delta k_{x}\approx $ $\Delta
k_{x0}+\delta k_{y}^{4}/uE$, $1+\chi _{\mathrm{I}_{1}}^{\ast }\chi _{%
\mathrm{I}_{2}}\approx -2i\delta k_{y}^{2}/E$, and $\chi _{\mathrm{I}%
_{1}}-\chi _{\mathrm{II}_{1}}\approx \lbrack 1/E-1/(E-U_{0})]\delta k_{y}^{2}
$. Therefore, the relevant parameters are $\lambda _{x}=4$, $\gamma _{x}=2$ and $c_{x0}=-2i/E$.
To obtain the parameter $\beta _{x}$, we need to calculate the reflectivity
amplitude. For a given $E$ and $k_{y}$, using the
continuity condition of the wave function, the reflectivity amplitude is
obtained as
\begin{equation}
r=-i\frac{2(\chi _{\mathrm{I}_{1}}-\chi _{\mathrm{II}_{1}})(\chi _{\mathrm{I}%
_{1}}-\chi _{\mathrm{II}_{1}}^{\ast })\mathrm{sin}(q_{x}d)}{|\chi _{\mathrm{I%
}_{1}}-\chi _{\mathrm{II}_{1}}|^{2}e^{iq_{x}d}-|\chi _{\mathrm{I}_{1}}-\chi
_{\mathrm{II}_{1}}^{\ast }|^{2}e^{-iq_{x}d}},
\end{equation}%
where $k_{x}$=$\sqrt{E^{2}-k_{y}^{4}}/u$ and $q_{x}$=$\sqrt{%
(U_{0}-E)^{2}-k_{y}^{4}}/u$. Expanding the reflectivity amplitude near the
EPs, we have $r\approx r_{x0}\delta k_{y }^2$ with $r_{x0}=\sin(U_0d/u)e^{-iU_0d/u}E^{-1}$, giving $\beta_x
=2$. One can also obtain the parameter $\beta_x$ by fitting the reflectivity
amplitude as a polynomial of $\delta k_{y }$ numerically, which is easier
than calculating the reflectively amplitude analytically. Hereafter, we will use
numerical fitting to obtain this parameter. Then, the power-law decay index
is $\nu =-(\beta _{x}+\gamma _{x}+1)/\lambda _{x}=-5/4$.
Therefore, according to the analysis, the LDOS oscillation in this case can
be fitted as $\delta \rho \left( x,E\right)\simeq \rho_A\cos \left( -\frac{2E}{u}x+\phi
\right) x^{-\frac{5}{4}}$. Based on Eq. (11), the amplitude $\rho_A$ depends on $r_{x0}$, $c_{x0}$ and the forth derivative of $\Delta k_x(k_y)$ at $(E/u,0)$. Taking all the factors together, we find the amplitude is proportional to $E^{-3/4}$ which means the higher the incident energy, the smaller the amplitude. Unfortunately, we cannot directly observe it in Fig. 2 because different incident energy corresponds to different pair of EPs having different initial phase.
The blue dashed lines in Figs. 2(c-f) show that the fitted results
for the LDOS are in excellent agreement
with the numerical ones. The asymptotic lines $x^{-5/4}$ well describe
the asymptotic behavior of the LDOS, which clearly demonstrate that the
decay index is $-5/4$. However, the LDOS oscillations close to the line defect departure from the asymptotic behavior because the asymptotic region requires $x$$\gg$$|\Delta k_{x0}|^{-1}$=$u/2E$ \cite{Wang,Biswas2,Liuq}. This means that the higher the incident energy, the smaller the distance needed to manifest asymptotic behavior. We can also directly observe this feature in Figs. 2(c-f). More precisely, based on our calculation, we find the asymptotic behavior expressed by Eq. (11) works quite well if $x$ is one order larger than $(\Delta k_{x0})^{-1}$.

On the other hand, it is worth to point out that the decay index $-5/4$
here is not only different from the decay of $-3/2$ for isotropic Dirac materials \cite{Xue,Biswas1,Biswas2,Wang} but also the decay of $-1/2$ for conventional semimetals \cite{Crommie,Petersen,Biswas2}. This unique decay index
originates from the unique anisotropic band structure around SDP. In particular, compared
with conventional semimetals, there is Klein tunneling suppressing
the backscattering at normal incidence, promising a faster decay
than that of $-1/2$ \cite{Wang}. In contrast to isotropic Dirac systems, electrons are
more difficult to transmit the barrier due to the severer mismatch
of the wave vector at oblique incidence, resulting in a slower decay
behavior than that of $-3/2$. Noteworthily, from the derivation, the decay index $-5/4$
is independent on the incident energy, the barrier height or the band
parameters. Therefore, the decay index $-5/4$ can serve as a
fingerprint to characterize 2D semi-Dirac electrons.
\begin{figure}[tbp]
\center
\includegraphics[bb=168 40 555 592, width=3.4 in]{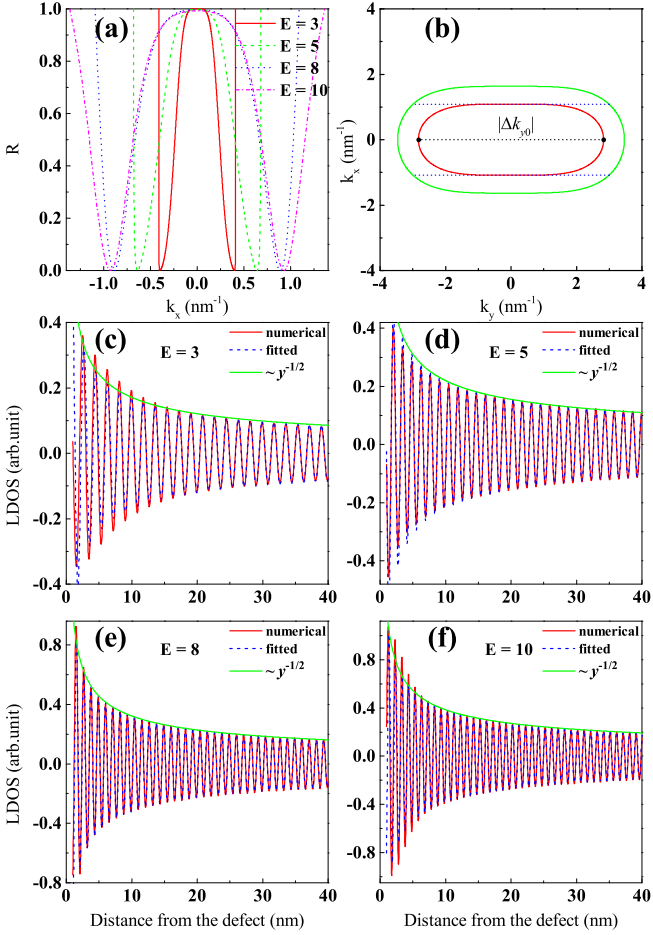} %
\renewcommand{\figurename}{FIG.}
\caption{(a) The reflectivity $R$ as a function of $k_x$ at different energies for $\protect\alpha =\protect\pi /2$. (b) The CECs of the incident (red line) and scattering (green line) regions
with $E=8$. The black solid dots are a pair of stationary phase points, and $%
|\Delta k_{y0}|$ is the characteristic wave vector during the scattering
process. (c)-(f) The LDOS oscillations at different energies. The red solid (blue dashed)
lines are the numerical (fitted) results. The green solid lines are the envelop function of
LDOS. Other
parameters are the same as those in figure 2.}
\end{figure}

Next, we turn to another special case of $\alpha=\pi/2$. Fig. 3(a) shows the
reflectivity $R$ as a function of the transverse wave vector $k_x$ with various energies. In
contrast to the case of $\alpha=0$, the reflectivity here is no longer zero
at normal incidence i.e., $k_x=0$, due to the absence of Klein tunneling. Instead, there is
a total reflection at normal incidence resulting from the server mismatch of
the wave vectors [see Eq. (4)] between the incident and scattering states
because of the high electrical barrier. This is consistent with the previous
results \cite{Banerjee2}. Similarly, Figs. 3(c-f) depict the LDOS
oscillations in this case with $E=3,5,8,10$, respectively. As shown in the
figure, numerical results (the solid red lines) indicate that the LDOS
periodically oscillates with distance away from the defect with decreasing
amplitude, which also implies an asymptotic behavior in the oscillation
pattern. Fig. 3(b) plots the CEC of the incident and barrier regions,
respectively. From Fig. 3(b), we find a pair of EPs $(0,\sqrt{E})$
and $(0,-\sqrt{E})$ satisfying Eq. (9) (see the black solid dots), giving
the characteristic wave vector as $|\Delta k_{y0}|=2\sqrt{E}$. Hence, the
period of the LDOS oscillation in this case is $2\pi/|\Delta k_{y0}|=\pi/ \sqrt{E}$,
which also means the higher the energy, the faster the oscillation pattern. Those
features are well reflected in Figs. 3(c)-(f). Meanwhile, for a certain incident
energy, the period here is smaller than that in the case of $\alpha=0$.
Following the same process of $\alpha=0$, we can also obtain
the asymptotic behavior of the LDOS by using the stationary phase approximation. Specifically, near the EPs, we have $\Delta
k_{y}\approx $ $\Delta k_{y0}+u^{2}\delta k_{x}^{2}/(2E\sqrt{E})$, $1+\chi _{\mathrm{I}_{1}}^{\ast }\chi_{\mathrm{I}_{2}}\approx 2$. Hence, the relevant parameters are $\lambda_{y}=2$, $\gamma_{y}=0$, and $c_{y0}=2$. As plotted in Fig.
3(a), the reflectivity is almost unit around the EP,
giving the parameter $\beta_{y}=0$ and $r_{y0}=1$. Therefore, the
power-law decay index in this case is $\nu =-(\beta_{y}+\gamma_{y}
+1)/\lambda_{y}=-1/2$, which means the spatial dependence of the LDOS can be
expressed as $\delta \rho \left(y,E\right) \simeq \rho_A \cos \left( -2\sqrt{E}y+\phi \right)
y^{-\frac{1}{2}}$ with $\rho_A=\pi(\partial^2 \Delta k_y/\partial k_x^2)^{-1/2}/\sqrt{2}$. In this case, the amplitude $\rho_A$ only depends on the reciprocal of the square root of the second derivative ($\partial^2 \Delta k_y/\partial k_x^2=u^{2}/E^{3/2}$), which is just the square root of the curvature at EPs on the CEC.
The blue dashed lines in Figs. 3(c-f) show the results fitted with the above formula for the LDOS. As expected, the fitted results are in good agreement with the
numerical ones. And, the asymptotic line $y^{-1/2}$ well describes the
asymptotic behavior of the LDOS far away from the line defect. This decay index is the same as that in
conventional metals \cite{Crommie,Petersen,Biswas2} because both of them have a parabolic dispersion. In the very vicinity of the line defect, the LDOS oscillations departure from the asymptotic behavior because the asymptotic region requires $y\gg|\Delta k_{y0}|^{-1}=1/2\sqrt{E}$ \cite{Wang,Biswas2,Liuq}. This means that the higher the incident energy, the smaller the distance needed to manifest asymptotic behavior. We can also directly observe this feature in Figs. 3(c-f). Specifically, we find that the asymptotic behavior starts when $y$ is twenty times larger than $(\Delta k_{y0})^{-1}$ based on our calculation.

Next, we discuss the case of $0<\alpha<\pi/2$, which means that the line
defect is along an arbitrary direction. In contrast to the two special
cases, the decay indexes in this case sensitively depend on the incident
energy and the orientation of the defect i.e., the titled angle $\alpha$,
because the CECs are titled ellipses. Here, we choose the results for $%
\alpha=\pi/4$ as an example to illustrate this feature. Figs. 4(a)-(b) show
the LDOS oscillations near the defect for $E=5$ and 8, respectively. From
the figures, we find that the numerical results (the red solid lines) also
imply an asymptotic behavior in the oscillation pattern. Following the
analysis for the two special cases ($\alpha=0$ and $\pi/2$), we can also try to fit
the LDOS by using a similar formula like Eq. (11). The fitted results and
asymptotic lines are indicated by the blue dashed and green solid lines in
the figures. The decay indexes here are found to be -0.843 and -0.603
for $E=5$ and 8, respectively. We have also checked various cases with different
incident energies and $\alpha$, but do not present them here due to space
limitations. The LDOS oscillations for other angle $\alpha$ and incident
energies are similar to the results shown in Figs. 4(a)-(b). However, the
decay indexes are distinct from each other and sensitively depend on the
incident energy and $\alpha$. The reason is that the CECs are titled ellipses
as depicted in Fig. 4(c) and there is no such a strict stationary phase point on the CEC
satisfying Eq. (9). The red solid line in Fig. 4(d) plots the decay indexes for various titled
angle $\alpha$ with incident energy $E=8$ by fitting with the numerical
results. The result shows that the decay index sensitively depends on the
orientation of the defect ($\alpha$) and varies from $-5/4$ to $-1/2$.

\begin{figure}[tbp]
\center
\includegraphics[bb=151 162 545 542, width=3.4 in]{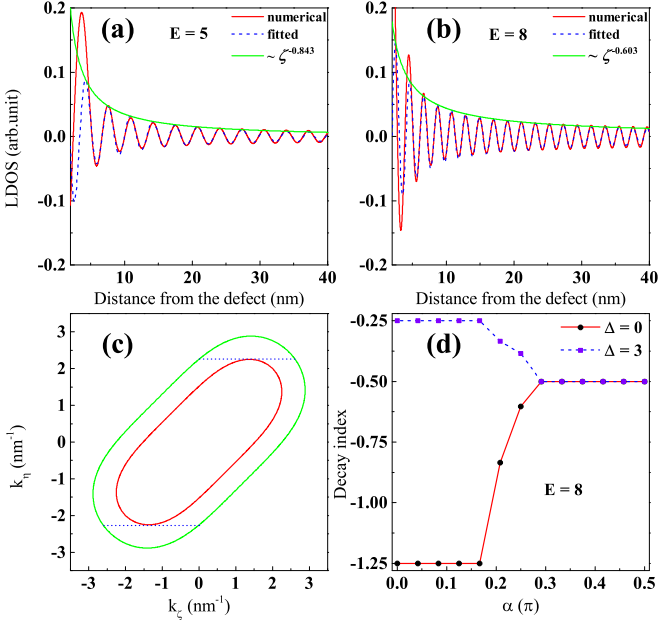} %
\renewcommand{\figurename}{FIG.}
\caption{(a)-(b) The LDOS oscillations at different energies for $\protect%
\alpha=\protect\pi/4$. (c) The CECs of the incident (red line) and scattering
(green line) regions with $E=8$. (d) The power-law decay index as a function the titled angle $\alpha$ with incident energy $E=8$. The red solid (blue dashed) line is the result for the system without (with) a gap. Other parameters are the same as those in
figure 2.}
\end{figure}

\begin{figure}[tbp]
\center
\includegraphics[bb=195 33 581 585, width=3.4 in]{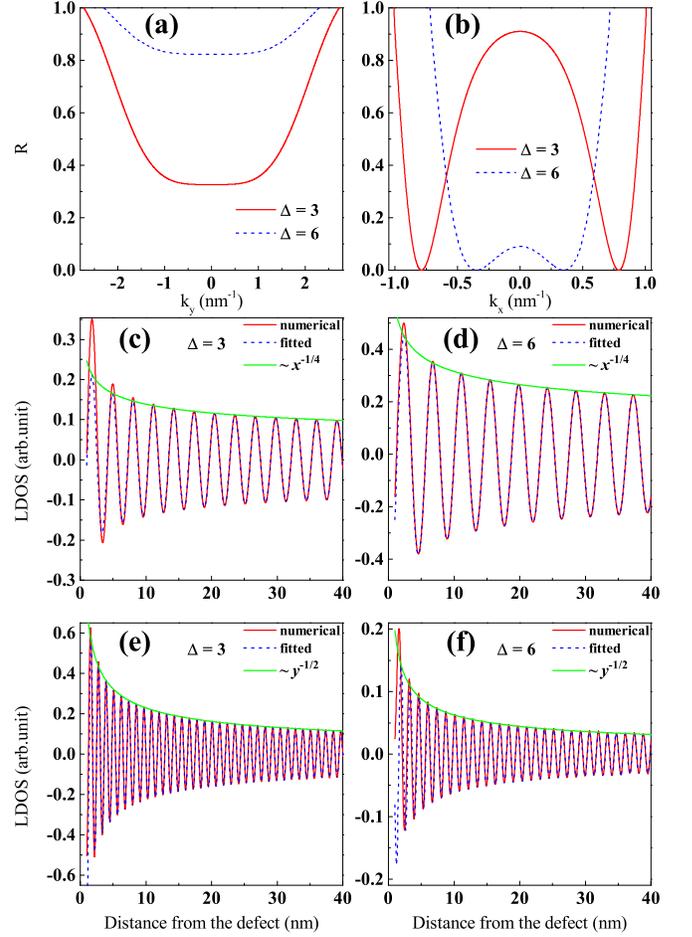} %
\renewcommand{\figurename}{FIG.}
\caption{(a)/(b) Reflectivity as a function of transverse wave vectors for $\protect%
\alpha=0$/$\frac{\protect\pi}{2}$ with different band gaps. (c)-(d) The
LDOS oscillations with different band gaps for $\protect%
\alpha=0$. (e)-(f) The
LDOS oscillations with different band gaps for $\protect%
\alpha=\pi/2$. The electron energy is $E=8$, and other
parameters are the same as those in figure 2.}
\end{figure}

To explore whether the decay index $-5/4$ is unique, we assume a gap (%
$\Delta $) in Hamiltonian (1) and redo the calculation. Figs. 5(a)-(b)
show the reflectivity as a function of transverse wave vectors for $\alpha=0$ and $%
\pi /2$, respectively. As shown in Fig. 5(a), for $\alpha =0$, the
reflectivity is finite at normal incidence due to the absence of Klein
tunneling resulting from the breaking of time reversal symmetry, which is different from the result of the gapless case. For $%
\alpha =\pi /2$, the reflectivity is nonzero at normal incidence regardless
of whether there is a band gap. This is similar to that of the gapless case.
Figs. 5(c)-(f) plot the LDOS oscillations for $\alpha=0$ and $\pi /2$ with different band gaps. Similar to the results of the gapless
case, the LDOSs oscillate periodically with a decay magnitude near the line
defect, which can be fitted by an analytical function as expressed in Eq.
(11). Here, for the case of $\alpha=0$, the stationary phase points on the
CEC are $(k_{x0}^{\prime},0)$ and $(-k_{x0}^{\prime},0)$ with $k_{x0}^{\prime}=\sqrt{%
E^{2}-\Delta ^{2}}/u$, giving the period of the LDOS oscillations as $\pi u/%
\sqrt{E^{2}-\Delta ^{2}}$. This means the larger the band gap, the slower the
oscillation pattern for a certain incident energy. Meanwhile, the asymptotic region falls into $x\gg u/2\sqrt{E^2-\Delta^2}$ which means the larger the band gap, the longer the distance needed to manifest asymptotic behavior for a certain energy. Those features can be
clearly observed in Figs. 5(c)-(d). Near those two points, we have $\Delta
k_{x}^{\prime}\approx -2k_{x0}^{\prime}+\delta k_{y}^{4}/u\sqrt{E^{2}-\Delta ^{2}}$ and $1+\chi _{%
\mathrm{I}_{1}}^{\ast }\chi _{\mathrm{I}_{2}}\approx 2\Delta /(E+\Delta )$.
Thus, the relevant parameters are $\lambda _{x}^{\prime }=4$, $\gamma
_{x}^{\prime }=0$, and $c_{x0}^{\prime}$=$2\Delta /(E+\Delta)$. The reflectivity amplitude around $k_{y}=0$ is $r\approx r_{x0}^{\prime}$ with $r_{x0}^{\prime}$=$U_0\Delta/[2E(E-U_0+\Delta)-U_0\Delta+i\sqrt{uq_x^{\prime}(E^2-\Delta^2)}\cot(q_x^{\prime}d)]$ and $q_x^{\prime}=\sqrt{(U_0-E)^2-\Delta^2}/u$, giving the parameter $\beta _{x}^{\prime }=0$ which is different from the result
of the gapless case. Therefore, the power-law decay index for gapped
semi-Dirac system is $\nu =-(\beta _{x}^{\prime }+\gamma _{x}^{\prime
}+1)/\lambda _{x}^{\prime }=-1/4$. The LDOS oscillations can be fitted as $\delta\rho\left( x,E\right) \simeq\rho_{Ax}^{\prime}\cos \left( -2k_{x0}^{\prime}x+\phi \right)
x^{-\frac{1}{4}}$. The amplitude $\rho_{Ax}^{\prime}$ depends on $r_{x0}^{\prime}$, $c_{x0}^{\prime}$ and the forth derivative of $\Delta k_x^{\prime}(k_y)$ at $(k_{x0}^{\prime},0)$. It is a complex function of the incident energy and band gap.
The blue dashed lines in Figs. 5(c-d) show the fitted results for the LDOS. They are in good agreement with the
numerical ones. Moreover, the envelop function $x^{-1/4}$ perfectly describes the
asymptotic behavior of the LDOS. It is worth to point out that the $-1/4$ decay behavior of the LDOS is also valid near the bottom (top) of the conduction (valence) band
for small band gaps. For higher energy LDOS oscillations with small gap, the decay index may be different, but it is usually difficult to realize in real experiments because higher Fermi surface requires high carrier density in the sample. On all accounts, the decay index can't be $-5/4$ (the decay index of gapless case) as long as the SDP is gapped, which further indicates that the $-5/4$ decay behavior can serve as a fingerprint to verify the SDP in 2D electron system.
Similarly,
for the case of $\alpha=\pi/2$, the extreme points are $(0,k_{y0}^{\prime})$
and $(0,-k_{y0}^{\prime})$ with $k_{y0}^{\prime}=(E^{2}-\Delta^{2})^{1/4}$. Near those two points, we have $\Delta k_{y}^{\prime}\approx
-2k_{y0}^{\prime}+u^{2}\delta k_{x}^{2}/2(E^{2}-\Delta^{2})^{3/4}$, and $1+\chi _{\mathrm{I}%
_{1}}^{\ast }\chi_{\mathrm{I}_{2}}\approx 2E/(E+\Delta)$, which results in the relevant parameters $\lambda^{\prime
}_{y}$=2 and $\gamma^{\prime }_{y}$=0. It can be seen from Fig. 5(b) that the reflection amplitude is not zero near the extreme points, thus, we obtain $\beta^{\prime }_{y}=0$. Hence, the decay
index in this case is $\nu =-(\beta _{y}^{\prime }+\gamma _{y}^{\prime
}+1)/\lambda _{y}^{\prime }=-1/2$, which is same as that of the gapless case. Therefore, the LDOS oscillation can be fitted as $\delta \rho (y)\propto \cos
\left(-2k_{y0}^{\prime}y+\phi\right)y^{-\frac12}$. The amplitude of the LDOS in this case is a complex function of the incident energy and band gap. It is more easily to obtain it by fitting with the numerical data. The blue dashed lines in Figs. 5(e)-(f)
are the fitted results, which are in good agreement
with the numerical results. The envelop function $y^{-1/2}$ perfectly governs the asymptotical behavior of the LDOS oscillations, which indicates that the power-law decay index is $-1/2$ for defects perpendicular to the parabolic dispersion direction regardless of whether the SDP is gapped or not.
The asymptotic region in this case is $x\gg (E^2-\Delta^2)^{-1/4}/2$ which also indicates that the larger the band gap, the longer the distance needed to manifest asymptotic behavior for certain energies. For $0<\alpha<\pi/2$, the decay indexes also depend on the incident energy and the orientation of the defect. The blue dashed line in Fig. 4(d) plots the decay indexes for various titled
angle $\alpha$ with incident energy $E=8$ and gap $\Delta=3$ by fitting with the numerical
results. From the figure, we find that the decay index depends on the
orientation of the defect and varies from $-1/4$ to $-1/2$.

\section{Summary}

In summary, using quantum mechanical scattering theory and the method of
stationary phase approximation, we studied the LDOS oscillations near a line defect in the semi-Dirac electron system and analytically obtained the power-law decay indexes for two special orientations of the defect. When the line defect is perpendicular to the linear dispersion direction, the decay index is $-5/4$ for gapless SDP
and $-1/4$ when the SDP is gapped, both of them are different from the decay index $-3/2$ in
isotropic Dirac systems. When the line defect is perpendicular to the parabolic dispersion direction, the decay index is always $-1/2$ regardless of whether the SDP is gapped or not. This is the same as that of conventional metals because both of them have a parabolic dispersion. There is no such a universal decay index when the line defect runs along an arbitrary direction due to the absence of stationary phase points on the CEC. Our results can be tested by the scanning tunnelling microscope \cite{Petersen,Crommie,kehuiwu,Avraham,Xue}, and the decay index $-5/4$ provides a fingerprint to detect semi-Dirac electrons. The power-law decay behavior is more likely to manifest itself at higher Fermi levels, which means it is more easily to be observed in samples with higher carrier concentrations.

\begin{acknowledgments}
This work was supported by the National Natural Science Foundation of China
(Grant Nos. 11804092 and 11774085), Project funded by China Postdoctoral
Science Foundation (Grant Nos. BX20180097, 2019M652777), and Hunan
Provincial Natural Science Foundation of China (Grant No. 2019JJ40187).
\end{acknowledgments}

\section*{Appendix}
\setcounter{equation}{0}
\renewcommand{\theequation}{A\arabic{equation}}
In this appendix, we discuss the LDOS oscillations near the line defect which is modeled as a $\delta$-potential.
When the line defects run along the linear dispersion direction, the potential profile is $U(x)=U_0\delta(x)$, which is the limitation of the rectangular potential $U(x)=U_0\Theta(x)\Theta(d-x)$ if $U_0$$\rightarrow$$\infty$, $d$$\rightarrow$$0$, and $U_0d\equiv$ constant. Since the momentum is conserved in the $y$-direction, we can express the wavefunction as $\Psi_{\mathbf{k}}(x,y)=e^{ik_yy}\varphi(x)$. Because the secular equation $[H(-i\partial_x,k_y)+U_0\delta(x)]\Psi_{\mathbf{k}}(x,y)=E\Psi_{\mathbf{k}}(x,y)$ is a first-order partial differential equation with respect to $x$, the wavefunction is discontinuous at $x=0$. The boundary condition at $x=0$ is given by \cite{Korniyenko,Shao}
\begin{equation}
\Psi_{\mathrm{k}}|_{x=0^+}=e^{-i\sigma_y\tau}\Psi_{\mathrm{k}}|_{x=0^-},
\end{equation}
where $\tau=U_0d/u$ is a constant. By using Eq. (A1), we obtain the reflection amplitude as
\begin{equation}
r=\frac{(k_y^4+iuk_xk_y^2)\sin\tau}{iuk_xE\cos\tau-E^2\sin\tau}.
\end{equation}
\begin{figure*}
\center
\includegraphics[bb=30 169 780 547, width=17 cm]{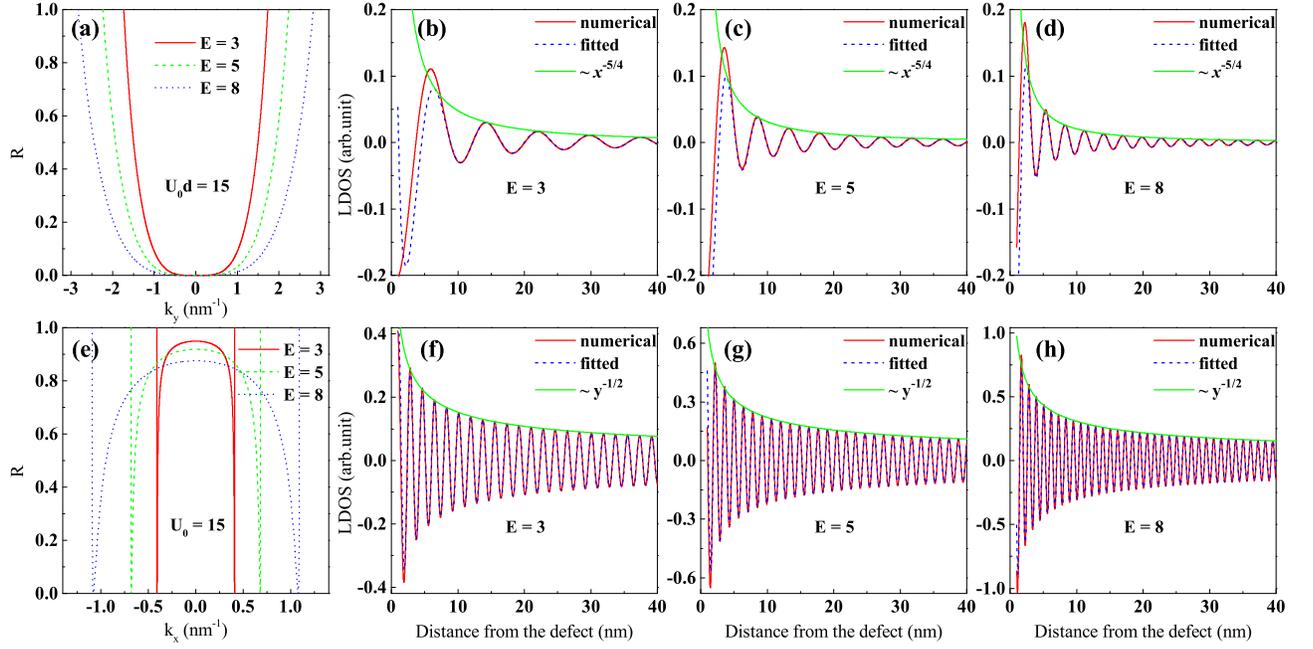} \renewcommand{%
\figurename}{FIG}
\caption{The first (second) row is the result for the defect perpendicular to the linear (parabolic) dispersion direction. (a)/(e) The reflectivity $R$ as a function of $k_y$/$k_x$ at different energies. (b)-(d) and (f)-(h) The spatial dependence of LDOS at different energies. The red solid (blue dashed) lines are the numerical (fitted) results. The green solid lines are the envelop function of
LDOS. The strength of the $\delta$ potential is $U_0d=15$ ($U_0=15$) when the line defect is perpendicular to the linear (parabolic) dispersion direction.}
\end{figure*}
Following the procedure of stationary phase approximation, expanding the reflectivity amplitude near the EPs $(\pm E/u,0)$ gives $r=\delta k_y^2\sin\tau/(Ee^{i\tau})$ which means the parameter $\beta_x$=2 which is the same as that in the case of rectangular potential. The other two relevant parameters $\lambda _{x}$ and $\gamma _{x}$ are solely determined by the CEC and independent on the choice of potential, i.e., $\lambda _{x}=4$, $\gamma _{x}=2$. Therefore, the power-law decay index here remains $\nu =-(\beta _{x}+\gamma _{x}+1)/\lambda _{x}=-5/4$ which is the same as that in the main text. Figs. 6(a) and 6(b-d) plot the reflectivity $R=|r|^2$ and LDOS oscillations counterpart to Figs. 2(a) and 2(c-e). As shown in the figures, we find there is only a little bit quantitative difference between the reflectivity and LDOS oscillations caused by the $\delta$ potential and the rectangular one. The numerical (the red solid lines) and fitted (the blue dashed lines) results both clearly demonstrate that the power-law decay index of LDOS oscillations are identical, i.e., decaying as $x^{-5/4}$.

When the line defect runs along the parabolic dispersion direction, the potential profile is $U(y)$=$U_0\delta(y)$. In this case, the wave function admits the form $\Psi_{\mathbf{k}}(x,y)=e^{ik_xx}\varphi(y)$. Because the Schr\"{o}dinger equation $[H(-i\partial_y,k_x)+U_0\delta(y)]\Psi_{\mathbf{k}}(x,y)$=$E\Psi_{\mathbf{k}}(x,y)$ is a second-order partial differential equation with respect to $y$, the boundary condition is \cite{Ting}
\begin{equation}
\begin{aligned}
\Psi_{\mathrm{k}}|_{y=0^-}=\Psi_{\mathrm{k}}|_{y=0^+},\\
\sigma_x\left(\partial_y\Psi_{\mathrm{k}}|_{y=0^+}-\partial_y\Psi_{\mathrm{k}}|_{y=0^-}\right)&=U_0\Psi_{\mathrm{k}}|_{y=0}.
\end{aligned}
\end{equation}
According to Eq. (A3), the reflection amplitude is obtained as
\begin{equation}
r=\frac{U_0^2k_y-2U_0E}{2i(U_0E-2k_y^3)-(U_0^2k_y-2U_0E)}.
\end{equation}
Similarly, expanding the reflectivity amplitude near the pair of extreme points $(0,\pm\sqrt{E})$ gives $r=U_0/(2i\sqrt{E}-U_0)$ which means the parameter $\beta_y$=0. It is the same as that in the case of rectangular potential. The other two relevant parameters $\lambda _{y}$ and $\gamma _{y}$ are solely determined by the CEC and independent on the choice of potential profile. Therefore, the power-law decay index here remains $-1/2$ which is identical to that in the case of rectangular potential. Figs. 6(e) and 6(f-h) plot the reflectivity $R=|r|^2$ and LDOS oscillations counterpart to Figs. 3(a) and 3(c-e). From the figures, we find there is only a little bit quantitative difference between the reflectivity and LDOS oscillations caused by the $\delta$ potential and the rectangular one. The numerical (the red solid lines) and fitted (the blue dashed lines) results both clearly demonstrate that the power-law decay index of LDOS oscillations are identical, i.e., decaying as $y^{-1/2}$.

In summary, different choices of the potentials will not bring about different decay indexes. The reason is that the quasiparticle interference pattern only depends on the energy dispersion, i.e., the CEC of the system. The scattering potentials can not change the geometry of the Fermi surface or the stationary phase points. Hence, the power-law decay indexes of the LDOS oscillations caused by the line defect are independent on whether choose a $\delta$-potential or a rectangular one to model it.

\end{document}